\documentclass[11pt]{article}
\let\section=\subsection     \let\subsection=\subsubsection                
\usepackage{graphicx}

\usepackage[english]{babel}
\usepackage{amsmath} 

\newcommand{\funcd}[2]{\frac{\delta #1}{\delta #2}}
\newcommand{\ii}{\ensuremath{\mathrm{i}}}
\def\text#1{\mathrm{#1}}\def\im{\mathrm{Im}}
\def\re{\mathrm{Re}}
\newcommand{\funcint}[2]{\int{\mathrm{d}(#2)}{#1}}

\date{February 27, 2002}

\sf

\begin{document}
\begin{center}
  {\large \bf Renormalization of self-consistent Schwinger-Dyson
    equations\\ at finite temperature\footnote{Contribution to the Workshop
      on Gross properties of Nuclei end Nuclear Excitations XXX: ``Ultra
      Relativistic Heavy-ion Collisions'', Hirschegg,
      Jan. 14 -- 20, 2002}} \\[5mm]
  {H. van Hees$^{1}$, and J.  Knoll$^{2}$} \\[3mm]
  {\small \it $^1$ Universit\"at Bielefeld, Universit\"atsstra\ss{}e,
    D-33615
    Bielefeld\\
    $^2$ Gesellschaft f\"ur Schwerionenforschung (GSI) \\
    Postfach 110552, D-64220 Darmstadt, Germany\\[3mm]}
\end{center}
\begin{abstract}\noindent
  We show that Dyson resummation schemes based on Baym's $\Phi$-derivable
  approximations can be renormalized with counter term structures solely
  defined on the vacuum level. First applications to the self-consistent
  solution of the sunset self-energy in $\phi^4$-theory are presented.
\end{abstract}

For the description of physical systems with strong interactions generally
perturbative methods are insufficient. Rather on the basis of effective
field theories non-perturbative methods such as partial resummation schemes
have to be applied.  Thus one works in terms of non-perturbative
``dressed'' propagators. One of the prominent self-consistent schemes is
the so called $\Phi$-derivable approximation \cite{lw60,leeyang61,bk61} or
effective action formalism \cite{cjt74}. There the self-energy of the
dressed propagator $G$ is generated from a functional $\Phi$ by means of a
variational principle \vspace{-1mm}
\begin{equation}
\label{1}
\Sigma_{12}=2 \ii \funcd{\Phi[G]}{G_{12}},
\vspace{-1mm}
\end{equation}
where $\Phi$ is given by a truncated set of closed 2-particle irreducible
(2PI) diagrams which cannot be split by cutting two lines.  Hence $\Phi$ is
the generating functional for \emph{skeleton diagrams} and thus avoids
double counting problems. It was shown by Baym \cite{baym62} that for so
defined approximations the expectation values of Noether charges are
exactly conserved, especially those which arise from space-time symmetries
(i.e., energy, momentum, angular momentum), while Ward Takahashi identities
of external symmetries and even crossing symmetry for the self-energy and
higher vertex functions are violated \cite{baymgrin}.

For long time the question could not be settled, to which extent
self-consistent approximations of this kind can be renormalized by
\emph{temperature independent counter terms}. The problem is to isolate the
divergent vacuum structure which is inherent and mostly hidden in the
self-consistent self-energy due to the implied partial resummation. A
further complication is that these hidden sub-divergences can be
overlapping. Therefore it was necessary to analyze self-consistent partial
Dyson resummation schemes defined by a set of basic generating self-energy
skeleton diagrams with dressed propagators. Here skeleton diagrams are void
of any self-energy insertion in any of its lines. This is guaranteed by the
above mentioned 2PI property of the closed diagrams defining $\Phi[G]$.

In terms of perturbative diagrams this leads to an infinite iterative
insertion of all basic 1PI-diagrams. The sum of all these perturbative
diagrams defines the self-consistent self-energy which determines the
dressed propagator. Typical 1PI diagrams defining the self-consistent
vacuum self-energy and the temperature or matter dependent pieces in
$\phi^4$ theory up to the self-consistent sunset diagram are the following
\begin{equation}
\label{Sigma-diag}
-\ii \Sigma =
\underbrace{\raisebox{-0.43mm}{\parbox{38mm}{
      \includegraphics{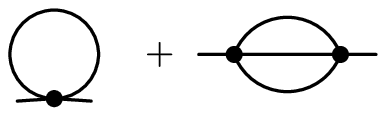}}}}_{\text{\mbox{\footnotesize basic
      diagrams}}}\; + 
 \cdots + \underbrace{\raisebox{4mm}{\parbox{57mm}{
       \includegraphics{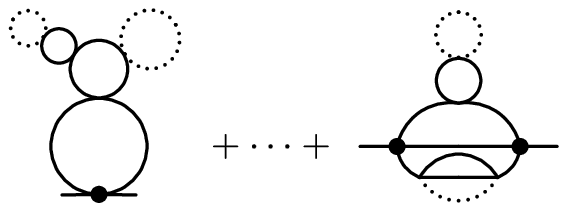}}}}_{\text{\mbox{\footnotesize
       generated perturbative diagrams}}}
\end{equation}
Here the full lines specify the self-consistent vacuum propagators, while
the dotted lines denote its temperature dependent pieces given by the KMS
condition, i.e., which contain the Bose-Einstein factors.  The basic
diagrams define the self-consistent vacuum self-energy, which needs to be
renormalized. At finite temperature the generated diagrams appear,
resulting from iterative insertions of basic type diagrams,
replacing one or more lines by dotted lines and omitting all diagrams which
contain vacuum self energy insertions (perturbation theory starting from
the self consistent vacuum level). Also in these $T$-dependent diagrams
\emph{all} subdiagrams entirely built up by vacuum propagators, which have
at most four external lines, are divergent and have to be renormalized.
These structures, however, appear in a nested and overlapping way
such that first the innermost subdiagrams have to be renormalized
through counter-terms given by the reduced diagrams where the divergent
sub-pieces are contracted to a point. The so obtained reduced diagrams
themselves are to be subjected to the same procedure. This iterative
process is formalized as the BPHZ-renormalization scheme. For the
self-consistent scheme under consideration the key issue is to find a
compact iteration procedure that generates all the required counter-terms
at once.

Within the real-time formalism \cite{Sch61,kel64} we could show for the
first time in general terms that such a strategy is possible
\cite{vHK2001-Ren-I}. In order to isolate the divergent vacuum subdiagrams
from the temperature dependent remainder the use of the real-time
formulation and the BPHZ-prescription of renormalization theory is crucial:
By a systematic expansion around the self-consistent vacuum solution we
split the self-energy and the propagator in three parts
\begin{equation}
\label{sig-split}
\Sigma=\Sigma^{(\text{vac})}+\Sigma^{(0)}+\Sigma^{(r)}, \; G=G^{(\text{vac})}
+ G^{(\text{vac})} \Sigma^{(0)} G^{(\text{vac})} + G^{(r)}.
\end{equation}
Here the vacuum parts are the renormalized self-consistent \emph{diagonal
elements} of the real-time matrix $\{-+\}$-formalism for the two-point
functions, while $\Sigma^{(0)}$ is the self-energy part which is linear in
$G-G^{(\text{vac})}$, i.e.,
\begin{equation}
\Sigma^{(0)}=\funcint{\Gamma^{(4,\text{vac})}_{12;34} [G-G^{(\text{vac})}]_{34}}{34}
\quad\text{\mbox{ with }} \quad
\Gamma_{12;34}^{(4,\text{vac})}= \left . -2\ii \frac{\delta^2
    \Gamma[G]}{\delta G_{12} 
    \delta G_{34}} \right|_{T=0}.
\end{equation}
The abbreviation $\funcint{}{1\ldots k}$ stands for the integration over
the space-time variables $1, \ldots, k$ with the time variable running
along the real-time Schwinger-Keldysh contour. Power counting together with
Weinberg's theorem shows that the asymptotic behavior of the various
self-energy parts in (\ref{sig-split}) is $O(p^{2})$, $O(p^0)$, and
$O(p^{-2})$ (modulo logarithms which are unimportant concerning the
convergence properties of the self-consistent diagrams) respectively. Since
$\Sigma^{(0)}$ is linear in $G-G^{(\text{vac})}$ it is clear that it obeys
itself a linear integral equation in terms of $G^{(r)}$ which behaves
asymptotically as $O(p^{-6})$:
\begin{equation}
\label{sig0-La}
\Sigma_{12}^{(0)} = \funcint{\Lambda_{12;34} G_{34}^{(r)}}{34},
\end{equation}
where $\Lambda$ has to be determined by the \emph{vacuum Bethe-Salpeter
  (BS) ladder-equation}
\begin{equation}
\Lambda_{12;34} = \Gamma_{12;34}^{(4,\text{vac})} + \ii
\funcint{\Gamma_{12;3'4'}^{(4,\text{vac})} G^{(\text{vac})}_{3'5}
  G^{(\text{vac})}_{4'6} \Lambda_{56;34}}{3'4'56}.  
\end{equation}
The corresponding momentum integral is logarithmically divergent and a
detailed BPHZ-analysis yields a set of renormalized equations in momentum
space \def\di{\mathrm{d}}
\begin{align}
\label{diff1-ren}
\Lambda^{(\text{ren})}(0,q)=&\Lambda^{(\text{ren})}(0,0) + 
\Gamma^{(4,\text{vac})}(0,q) -\Gamma^{(4,\text{vac})}(0,0)  \\
\nonumber
& + \ii \int\frac{\di^4 l}{(2\pi)^4}
\Lambda^{(\text{ren})}(0,l)[G^{(\text{vac})}(l)]^2 
[\Gamma^{(4,\text{vac})}(l,q) -\Gamma^{(4,\text{vac})}(l,0)],\\
\label{diff2-ren}
\Lambda^{(\text{ren})}(p,q)=&\Lambda^{(\text{ren})}(0,q) + 
\Gamma^{(4,\text{vac})}(p,q) -\Gamma^{(4,\text{vac})}(0,q)  \\
\nonumber
& + \ii \int\frac{\di^4 l}{(2\pi)^4}
[\Gamma^{(4,\text{vac})}(p,l) -\Gamma^{(4,\text{vac})}(0,l)]
[G^{(\text{vac})}(l)]^2 \Lambda^{(\text{ren})}(l,q),
\end{align}
to be solved in kind of sweep up (\ref{diff1-ren}) - sweep down
(\ref{diff2-ren}) method. Here $\Lambda^{\text{ren}}(0,0)$ defines the
renormalization condition. The 2PI property of $\Gamma^{(4,\text{vac})}$
and Weinberg's power counting theorem ensure that the differences of
$\Gamma^{(4,\text{vac})}$ appearing in the latter equations are of negative
degree of divergence with respect to their integration variable and thus
all these integrals are finite. In this pure vacuum equation all quantities
are to be read as the time ordered (i.e., the $\{--\}$-components). Since
$\Lambda^{(4,\text{vac})}(p,q)$ is of $O(q^0)$ for fixed $p$ the equation
(\ref{sig0-La}) is finite since $G^{(r)}(p)=O(p^{-6})$. The finite
temperature self-energy is thus rendered finite with \emph{pure vacuum
  counter terms} by substitution of the renormalized four-point BS-function
in (\ref{sig0-La}). The remainder $\Sigma^{(r)}$ is finite by itself after
subtracting the explicit vacuum subdivergences, since it contains at least
two factors $G-G^{(\text{vac})}$.

Furthermore we could show that also the generating functional can be
renormalized by the same counter-terms and finally the renormalized
self-energy is defined by Eq. (\ref{1}), where $\Phi$ is to be read as the
\emph{renormalized functional}.  An analytic continuation from the
real-time to imaginary-time quantities then provides renormalized
expressions for thermal equilibrium such as the thermodynamic potentials
(like pressure or entropy) which are thermodynamically and dynamically
consistent. For details of the complete renormalization formalism see
  \cite{vHK2001-Ren-I}.
\begin{footnotesize} \begin{figure}
\centering{
\begin{minipage}{0.43\textwidth}
\centerline{\includegraphics[width=0.95 \textwidth]{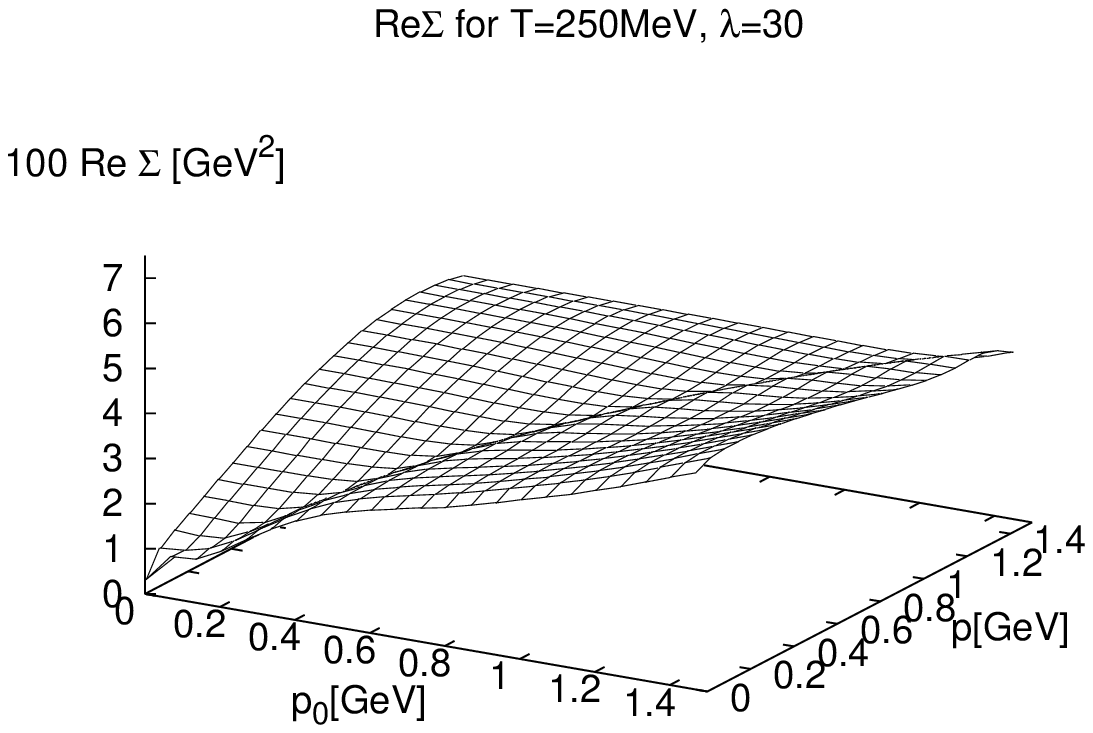}}
\end{minipage}\hfill
\begin{minipage}{0.43\textwidth}
  \centerline{\includegraphics[width=0.95
    \textwidth]{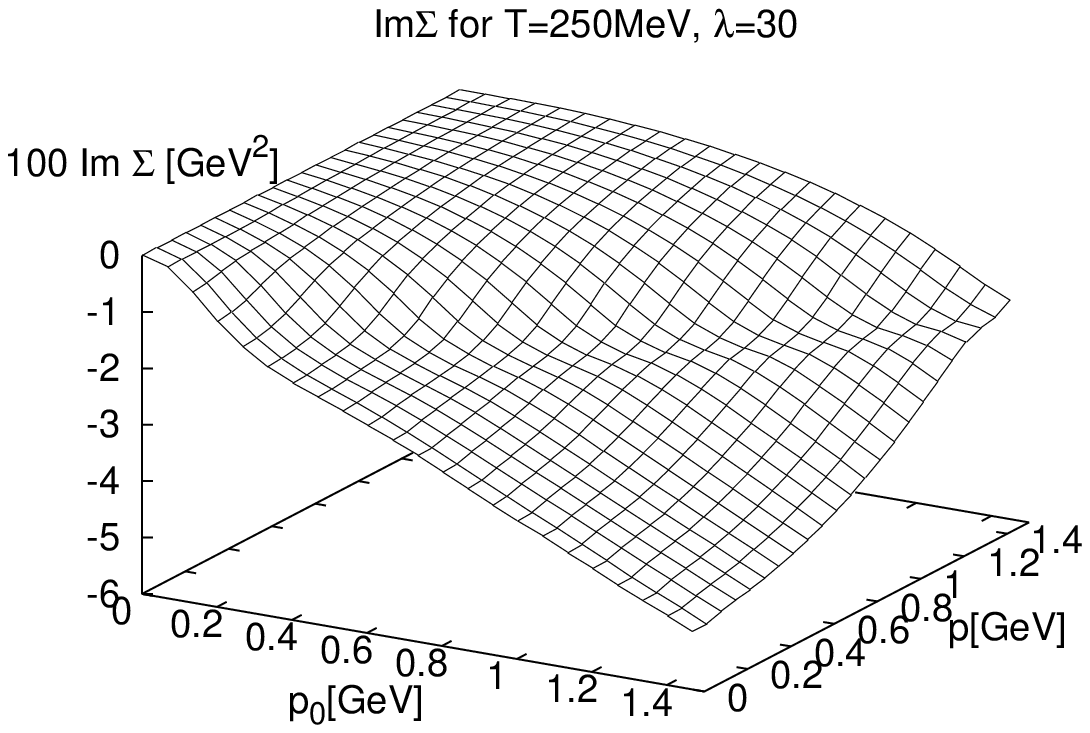}}
\end{minipage}
}
\caption{\footnotesize The real and imaginary part of the self-consistent self-energy for
  $\lambda/24 \phi^{4}$ were calculated. A full calculation without further
  approximations could be achieved. The main effect of self-consistency is
  that the higher masses due to the tadpole contribution, which is dominant
  for the real part of the self-energy lowers the phase-space available for
  decays into three particles while the growing finite width itself induces
  a further broadening.}
\label{fig1}
\end{figure}
\end{footnotesize}

First applications in next to leading order approximation to
$\phi^{4}$-theory, including both, the self-consistent tadpole and the
sunset-diagram for the self-energy, were presented
\cite{vHK2001-Ren-II}.  We could demonstrate that not only the
UV-problem but also the resulting singularities at the on-shell pole
of the vacuum propagator could be tackled numerically. The
self-consistent solutions are then obtained iteratively. The finite
temperature results for the self-consistent case are shown in Fig.
\ref{fig1} in a 3-dimensional plot over the $(p_0,|\vec{p}|)$-plane
illustrating that the entire calculations are performed with the full
dependence on energy and momentum on a $200 \times 200$ lattice.
Details can be extracted from the cuts shown for a set of selected
momenta in Fig.  \ref{fig5}.
\begin{footnotesize}
\begin{figure}
\centering{
\begin{minipage}{0.49\textwidth}
\centering{\includegraphics[width=\textwidth]{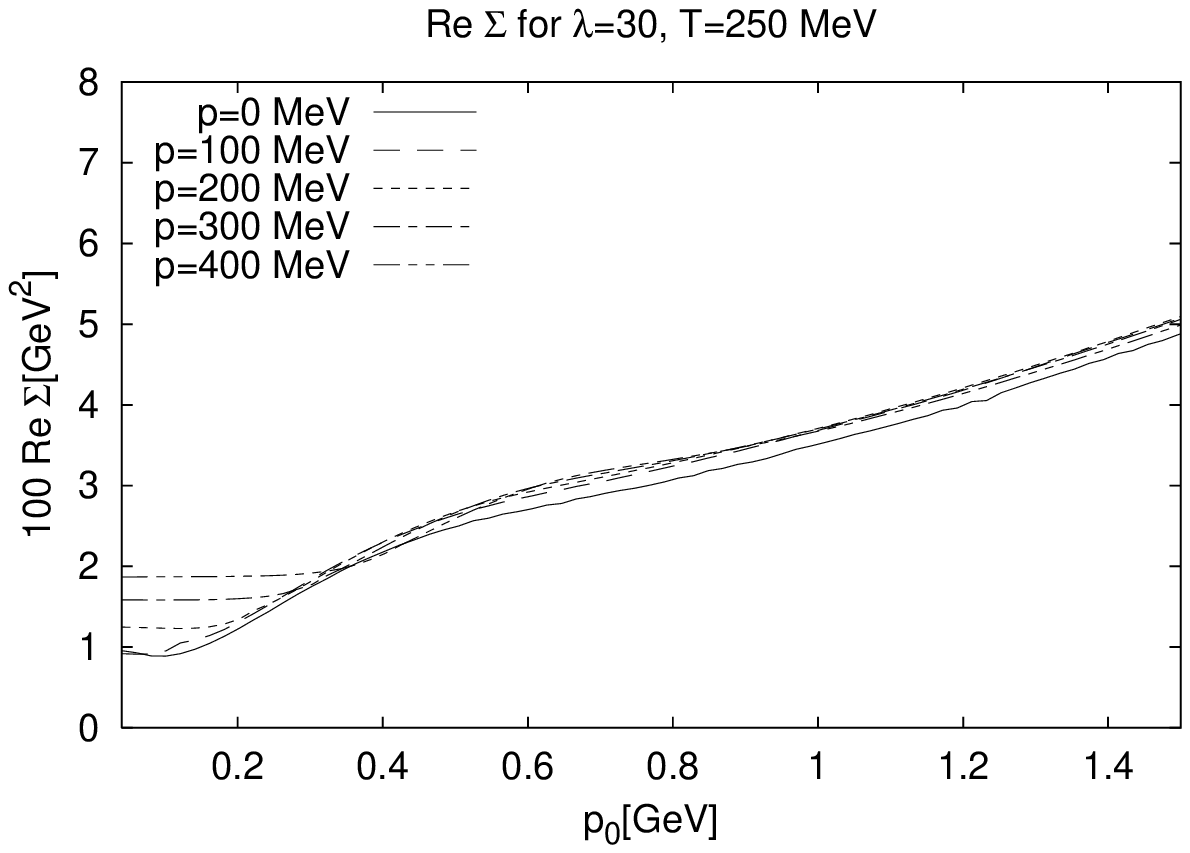}}
\end{minipage}\hspace*{0.2cm}
\begin{minipage}{0.49\textwidth}
  \centering{\includegraphics[width=\textwidth]{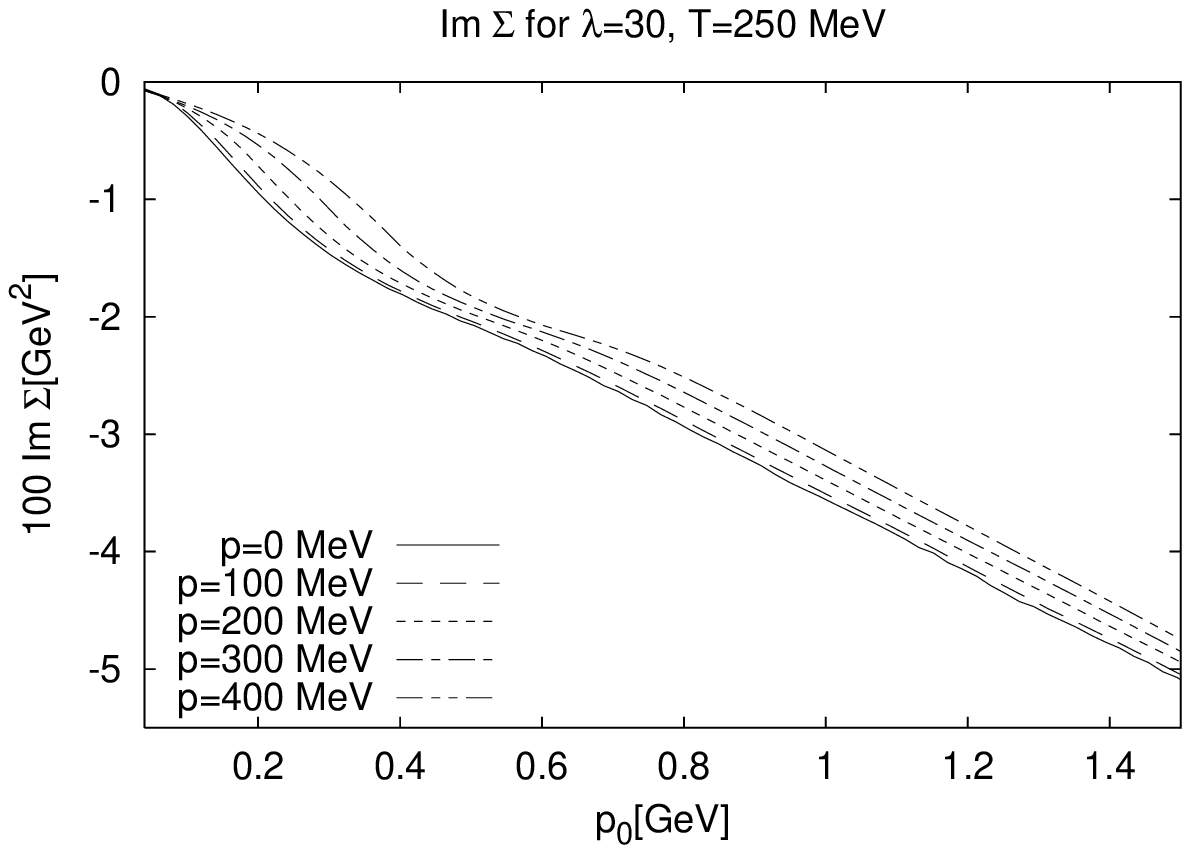}}
\end{minipage}
}
\caption{\footnotesize Real (left) and imaginary part (right) of the
  self-consistent 
  self-energy  for 
  $\lambda=30$, $m=140\text{MeV}$ and $T=250\,\text{MeV}$ as a
  function of $p_0$ for various 3-momenta.}
\label{fig5} 
\end{figure}\end{footnotesize}
The main qualitative results are similar for both the perturbative and the
self-consistent calculation: In the vacuum and self-consistent pure tadpole
case the self-energy shows a threshold cut resulting from the decay into
three particles, i.e., $p_0^2-\vec{p}^2\geq 9M^2$. Adding the sunset
self-energy leads to a spectral width which dissolves this threshold such
that the self-energy shows spectral strength (imaginary parts) at all
energies. While the growing high-energy tail is related to the decay of
virtual bosons into three particles, at finite temperature, as a new
component, a low-energy plateau in $\im \;\Sigma^{\text{R}}$ emerges from
in-medium scattering processes. 

Various balancing effects are encountered for the self-consistent case: For
sufficiently large couplings and/or temperatures the self-consistent
treatment shows quantitative effects on the width. The finite spectral
width itself leads to a further broadening of the width and a smoothing of
the structures as a function of energy.
\begin{figure}
\begin{footnotesize}
\centering{
\begin{minipage}{0.49\textwidth}
\centering{\includegraphics[width=\textwidth]{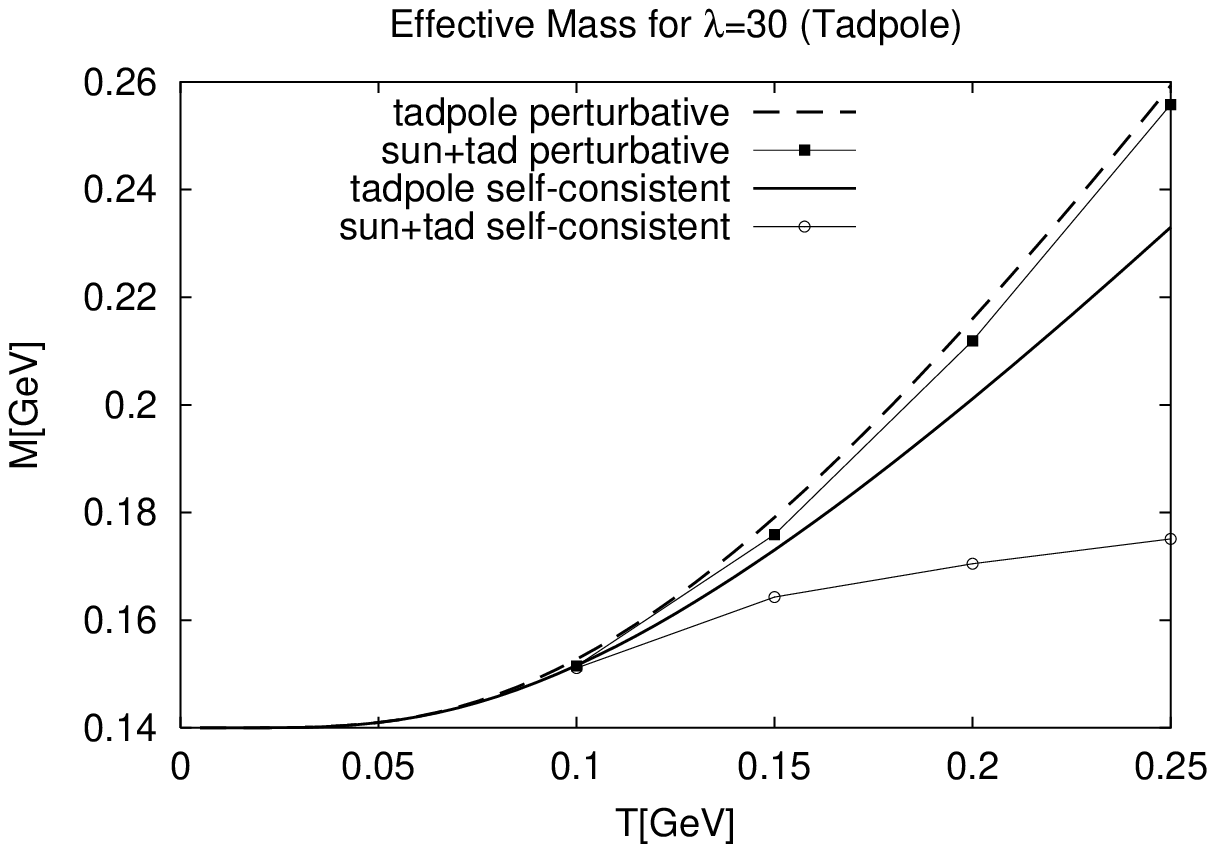}}
\end{minipage}\hspace*{0.2cm}
\begin{minipage}{0.49\textwidth}
\centering{\includegraphics[width=\textwidth]{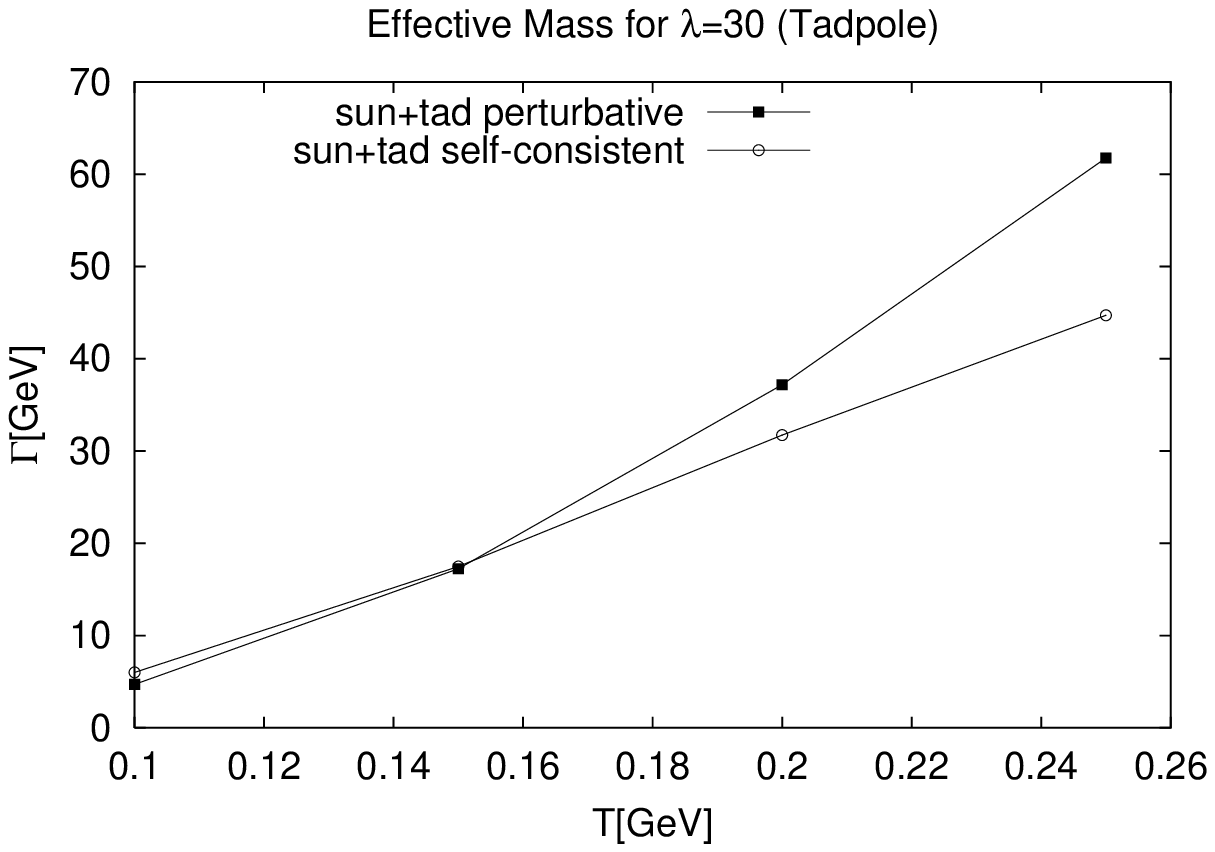}}
\end{minipage}}
\caption{\footnotesize The in-medium effective masses $M$ (left) and
  spectral widths 
  $\Gamma$ (right) of the particles for the various approximations
  described in the text as a function of the system's temperature $T$.}
\label{fig6}
\end{footnotesize}\end{figure}
This is however counter balanced by the behavior of the real part of the
self-energy, which, as discussed below, essentially shifts the in-medium
mass upwards. This reduces the available phase space for real processes.
With increasing coupling strength $\lambda$ a nearly linear behavior of
$\im \; \Sigma^{\text{R}}$ with $p_0$ results implying a nearly constant
damping width given by $-\im \; \Sigma^{\text{R}}/p_{0}$.

The overall normalization of the real part of $\Sigma^{\text{R}}$ is
determined by the renormalization procedure. In this case there are three
counter balancing effects. First the tadpole loop shifts the mass to higher
values. As the tadpole is less effective for higher masses this effect
weakens itself in the self-consistent tadpole treatment, c.f. Figs.
\ref{fig6}. However, since the sunset part adds spectral width, it
indirectly contributes to the tadpole loop. Since spectral strength at the
lower mass side carries higher statistical weights, the tadpole loop in
turn leads to a further increase of the mass shift, c.f.  the perturbative
calculations of sunset \& tadpole in Fig. \ref{fig6}.

The direct contributions of the sunset terms to the real part of the
self-energy become relevant at higher couplings and temperatures. Then the
self-consistency leads to significant effects which contributes to a net
down-shift of the real part of the self energy or in-medium mass $M$. The
latter effect finally overrules the tadpole shift and indeed leads to an
overall negative mass shift compared to the (tadpole dominated)
perturbative result.  These effects are illustrated in Fig.  \ref{fig6}
where the in-medium effective mass $M$ and width $\Gamma$ of the
corresponding ``quasi-particles'' are plotted against the temperature.
Thereby $M$ and $\Gamma$ are defined as the quasi-particle energy $M=p_0$
at vanishing real part of the dispersion relation $[p_0^{2}-m^2 -\re
\Sigma(p_{0},\vec{p})]_{p=(M,\vec{0})}=0$ for $\vec{p}=0$, and through
$\Gamma=-\im \; \Sigma(p)/p_{0}|_{p=(M,\vec{0})}$, respectively.

We have shown that it is possible to use the renormalization scheme,
proposed in \cite{vHK2001-Ren-I}, for numerical investigations of the
self-consistent approximations for the self-energy derived from the
truncated effective action formalism on the 2PI level. Thereby it is very
important to isolate the divergent vacuum parts consistently, in particular
the implicit or hidden ones, from all convergent and in particular
temperature or more generally matter-dependent parts. This could be
provided by the ansatz within the real-time formalism of quantum field
theory which allows to separate vacuum expressions from genuine
finite temperature parts of the propagator and the self-energy. The
results promise that the method, which is conserving \cite{baym62,kv97} and
thermodynamically consistent, can also be applied for the genuine
non-equilibrium case, i.e., in quantum transport \cite{KIV2001,IKV99} or
for the solution of the renormalized Kadanoff-Baym equations.

The investigation of the symmetry properties of $\Phi$-derivable
approximations is the subject of a forthcoming publication
\cite{vHK2001-Ren-III}. It is known that in general the symmetries of the
classical action which lead to Ward-Takahashi identities for the proper
vertex-functions are violated for the self-consistent Dyson resummation for
the functions beyond the one-point level, i.e., at the correlator level.
The reason is that, although the \emph{functional} $\Gamma$ can be expanded
with respect to expansion parameters like the coupling or $\hbar$ (loop
expansion) or large-$N$ expansions for O($N$) type models, the solution of
the self-consistent equations of motion contains partial contributions to
any order of the expansion parameter. This resummation is of course
incomplete and violates even crossing symmetry for the vertices involved in
the renormalization procedure. This causes problems concerning the
Nambu-Goldstone modes \cite{baymgrin} in the case of spontaneously broken
symmetry or concerning local gauge symmetries \cite{vHK2001-rho} when the
gauge fields are treated beyond the classical field level, i.e., at the
propagator level. In \cite{vHK2001-Ren-III} we discuss how to cure these
defects by supplementary vertex equations which remain renormalizable
following the strategy presented here. Such self-consistent
treatments are important for future applications to QCD or hadronic matter
problems at finite temperature and finite baryon densities. There further
complications arise due to the gauge structure of the gluon or vector
mesons \cite{vHK2001-rho}.  Further applications concern the derivation of
renormalized conserving quantum transport equations \cite{KIV2001} which
permit to treat broad resonances consistently \cite{IKV99}.

\begin{flushleft}

\end{flushleft}

\begin{thebibliography}{15}
\setlength{\itemsep}{-0.8mm}
\bibitem{lw60}
J.~Luttinger, J.~Ward, Phys. Rev. \textbf{118} (1960) 1417

\bibitem{leeyang61}
T.~D. Lee, C.~N. Yang, Phys. Rev. \textbf{117} (1961) 22

\bibitem{bk61}
G.~Baym, L.~Kadanoff, Phys. Rev. \textbf{124} (1961) 287

\bibitem{cjt74}
M.~Cornwall, R.~Jackiw, E.~Tomboulis, Phys. Rev. \textbf{D10} (1974) 2428

\bibitem{baym62}
G.~Baym, Phys. Rev. \textbf{127} (1962) 1391

\bibitem{baymgrin}
G.~Baym, G.~Grinstein, Phys. Rev. \textbf{D15} (1977) 2897

\bibitem{Sch61}
J.~Schwinger, J. Math. Phys \textbf{2} (1961) 407

\bibitem{kel64}
L.~Keldysh, ZhETF \textbf{47} (1964) 1515, [Sov. Phys JETP {\textbf{20}} 1965
  1018]

\bibitem{vHK2001-Ren-I}
H.~van Hees, J.~Knoll, Phys. Rev. \textbf{D65} (2002) 025010

\bibitem{vHK2001-Ren-III}
H.~van Hees, J.~Knoll, in preparation

\bibitem{vHK2001-Ren-II}
H.~van Hees, J.~Knoll, hep-ph/0111193, 
Phys. Rev. \textbf{D} in press

\bibitem{kv97} J. Knoll and D.N.Voskresenski,
Annals of Physics \textbf{249} (1996) 532

\bibitem{KIV2001} J. Knoll, Y.B. Ivanov and D.N.Voskresenski,
  Ann.Phys. (NY) \textbf{293} (2001) 126
\bibitem{IKV99} Y.B. Ivanov, J.Knoll and D.N.Voskresenski,
  Nucl.Phys.  \textbf{A 672} (2000) 313.

\bibitem{vHK2001-rho}
H.~van Hees, J.~Knoll, 
Nucl. Phys. \textbf{A683} (2001) 369

\end{thebibliography}
\end{document}